\documentclass[11pt,a4paper,superscriptaddress,showpacs]{revtex4}
\usepackage{color}
\usepackage{graphicx}
\usepackage{graphics}
\usepackage{amsmath}
\usepackage{dcolumn}
\usepackage{amssymb}
\usepackage{bm}
\usepackage[latin1]{inputenc}
\newcommand{\be}{\begin{equation}}
\newcommand{\ee}{\end{equation}}
\newcommand{\bea}{\begin{eqnarray}}
\newcommand{\eea}{\end{eqnarray}}

\begin{document}

\title{Effective potential for a SUSY Lee-Wick model: the Wess-Zumino case}

\author{M. Dias}
\email{mafd@cern.ch}
\affiliation{Departamento de Ci\^encias Exatas e da Terra,Universidade Federal de S\~ao Paulo\\
Diadema-SP-Brazil.}
\author{A. Yu. Petrov}
\email{petrov@fisica.ufpb.br}
\affiliation{Departamento de F\'{\i}sica, Universidade Federal da
Para\'{\i}ba\\
 Caixa Postal 5008, 58051-970, Jo\~ao Pessoa, Para\'{\i}ba, Brazil}
\author{C. R. Senise Jr.}
\email{carlossenise@unipampa.edu.br}
\affiliation{Universidade Federal do Pampa,\\
 Av. Pedro Anuncia\c{c}\~ao S/N, Vila Batista, 96570-000, Ca\c{c}apava do Sul, RS, Brazil}
\affiliation{Departamento de F\'{\i}sica Matematica, Instituto de
  F\'{\i}sica,
\\
Universidade de S\~ao Paulo,
 Caixa Postal 66318, S\~ao Paulo, Brazil} 
\author{A. J. da Silva}
\email{ajsilva@fma.if.usp.br}
\affiliation{Departamento de F\'{\i}sica Matematica, Instituto de
  F\'{\i}sica,
\\
Universidade de S\~ao Paulo,
 Caixa Postal 66318, S\~ao Paulo, Brazil}


\begin{abstract}
Using a superfield generalization of the tadpole method, we study the one-loop effective potential for a
Wess Zumino model modified by a higher-derivative term, inspired by the Lee-Wick model. The one-loop K\"{a}hlerian potential is also obtained by other methods and compared with the effective potential.

\end{abstract}

\maketitle

\section{Introduction}

The use of higher derivatives has been suggested long time ago as a way to obtain a better ultraviolet behavior for physically relevant models. In this context, in the early 1970s, Lee and Wick (LW) proposed a finite theory of QED \cite{1,2} and following that paper various improvements were done on that direction, in several different models. Recently, the interest in that theory was renewed in the so-called Lee-Wick Standard Model (LWSM), leading to new insights in the hierarchy problem and to some applications in phenomenology and cosmology \cite{Accioly:2010js,Espinosa:2011js,Karouby:2011wj,Carone1}. 

The LWSM assumes a minimal set of higher-derivative quadratic terms which produce a negative-norm partner for each Standard Model (SM) particle (some works also consider models with more than a single LW partner for each SM particle~\cite{Carone2}). On the other hand, a similar doubling of particles occurs for supersymmetric extensions of the SM: each SM particle gets a supersymmetric partner with the same gauge quantum numbers but different spin. 

Therefore, it is natural to discuss higher-derivative extensions of supersymmetric (SUSY) models, that is, to extend SUSY models by incorporating higher derivative terms (a similar theory was explored on \cite{galileon}, as an extension for the {\it galileon} field on superspace). In the past, within the superfield context, higher derivatives were introduced as a regularization tool \cite{Ili}. Further, it was shown that higher derivatives naturally emerge within the supergravity context, if one considers the supertrace anomaly of matter superfields in a curved superspace and then adds it to the classical action, generating higher derivative terms in the effective $N=1$ supergravity action \cite{sugra}. Some aspects of the dynamics of higher derivative superfield models have been studied in the works of Ref. \cite{aspects}. Besides that, since realistic models require that SUSY should be broken \cite{teste}, it is natural to study if a Lee-Wick supersymmetric extension  may exhibit some sort of dynamical SUSY breaking.

The introduction of higher-derivative terms is also justified when we look at the modification of the superficial degree of divergence, $\omega$, when compared with its usual counterpart \cite{Buchbinder:1996xf}. In the following, we outline the key ingredients to derive these modifications in $\omega$. Any integration over the internal momentum ($d^4p$) contributes to $\omega$ with a factor of 4. Since the number of integrations over internal momenta is equal to the number of loops, $L$, the total contribution from such integrations is $4L$. There are two types of internal lines, $L_1$ and $L_2$ (see (\ref{beer99}) below), referring respectively to $\Delta^{\bar{S}S}$ and  $\Delta^{S\bar{S}}$; they include, illustrating the case for $L_1$, a $\frac{1+p^2/M^2}{p^2(1+p^2/M^2)^2-m^2}\propto\frac{1}{p^4}$ factor, hence, its contribution is $-6L_1$, not forgetting an extra term $\frac{1}{p^2}$ coming from the definition of $P_1$ (see Eq.~(\ref{lc1})).

Now let us consider the $D-$factors: the $L_1$ propagators contain a factor of $D^2(\bar{D}^2)$; also four D-factors (each one with canonical dimension $1/2$) are consumed in the identities $\Phi=-\frac{1}{4} \bar D^2 S$ and $\bar \Phi=-\frac{1}{4} D^2 \bar S$, where $S$ and $\bar S$ are unconstrained scalar superfields in terms of which we write the $\Phi$ and $\bar\Phi$ superfields. All of these increase by $2L_1$ their contribution. By the same arguments, we find out a $-6L_2$ contribution to $\omega$. 

For the chiral (antichiral) $\Phi(\bar{\Phi})$ superfield, the Feynman rules dictate that each vertex, $V$, without external lines, includes four D-factors; only two of these are associated to a vertex in the presence of an external line, $E$. So we shall add a  $4V-2E$ term. We have to consider that each loop gives a local $\theta-$space contribution, and using 
\[\delta^{4}(\theta-\theta^{\prime})D^2\bar{D}^{\prime2}\delta^{4}(\theta-\theta^{\prime})=16\delta^{4}(\theta-\theta^{\prime}) \ ,\]
we can eliminate four D-factors to create a delta. We shall discount it including a factor of $-2L$. 

Summing it up, one obtains
\begin{equation} 
\label{LC1}
\omega\leq 4V-2E+2L -4L_1-6L_2 \ .
\end{equation}
 Nevertheless, in the $M\to \infty$ limit, we have, analogously
\begin{equation}
\label{LC2}
\omega\leq 4V-2E+2L-2L_1-2L_2 \ .
\end{equation}
Comparing (\ref{LC1}) and (\ref{LC2}) we note that a better behavior is achieved when one introduces higher-derivative terms with respect to the divergences in the original theory, considering the same number of $L_1$ and $L_2$ internal lines. With this motivation we calculate the effective potential for the Wess-Zumino model with an extra higher derivative term, by using the tadpole method  \cite{tadpole}. For that we must shift the fields by a classical superfield. We will also calculate the one-loop K\"{a}hlerian potential by another method and compare with the effective potential.

The work is structured as follows: in the next Section we shall study the modified Wess-Zumino model and calculate its one-loop effective potential. In Section III, we obtain the one-loop K\"{a}hlerian potential. In Section IV, we present our conclusions.
 

\section{One-Loop Effective Potential}

We start with the higher-derivative $N=1, D=4$ supersymmetric Lagrangian ($g_{\mu\nu}=diag(-+++)$), given by
\begin{equation}
\label{dante1}
{\mathcal S}=\int d^8z \left[\bar{\Phi}\Phi-\frac{1}{M^2}\bar{\Phi}\Box\Phi\right]+\left(\int d^6z\, W(\Phi)+h.c.\right) \ ,  
\end{equation} 
where $W(\Phi)=\frac{1}{2} m \Phi^2 +\frac{1}{6} g \Phi^3$ and $M$ is a mass parameter much bigger than $m$, but much smaller than the ultraviolet cut-off. By expanding the superfield in component fields and integrating in the grassmanian variables we obtain for the Lagrangian:
\begin{eqnarray}\label{rev1}\nonumber
{\cal L}_{WZ}&=&A^{\dagger}\Box\left(1-\frac{\Box}{M^2}\right)A+i\bar{\chi}\bar{\sigma}^\mu\left(1-\frac{\Box}{M^2}\right)\partial_\mu\chi+F^{\dagger}\left(1-\frac{\Box}{M^2}\right) F\\
&&+\left(\frac{\partial W}{\partial A} F-\frac{1}{2}\frac{\partial^2W}{\partial A^2}\chi\bar{\chi}+h.c.\right),
\end{eqnarray}
where $W(A)=W(\Phi=A)$, $\frac{\partial W}{\partial A}=mA+\frac{1}{2}g A^2$ and 
$\frac{\partial^2 W}{\partial A^2}=m+gA$. From this Lagrangian, we can derive the Euler-Lagrange equation for the ``auxiliary field'' $F$, e.g.  
\begin{eqnarray}\label{matan1}
\left(1-\frac{\Box}{M^2}\right)F&=&-\frac{\partial W^\dagger}{\partial A^\dagger} \ .
\end{eqnarray}
As can be seen, it now possesses dynamics. This kind of model has already been extensively studied in the literature, and it can be formulated as a second order theory by doubling the number of fields \cite{Antoniadis:2007xc}. For our purposes, i.e., for the evaluation of the effective potential, this formulation is not necessary, and we keep working with the original Lagrangian (\ref{dante1}).

The tree level potential, as usual, is given by the scalar part of the Lagrangian, with the scalar components of the chiral superfield set as constants (independent of $x$), $\Phi=\Phi_c=a+f \theta^2$:  
\begin{equation}
V_0=-ff^{\dag}-\left(ma+\frac{g}{2}a^{2}\right)f-\left(ma^{\dag}+\frac{g}{2}a^{\dag2}\right)f^{\dag} \ . \label{matan3}
\end{equation}

There is at least one solution for the minimum of $V_0$ with $f=0$. So, as in the usual Wess-Zumino model, SUSY is not broken at the tree level, and thus the nonrenormalization theorem dictates what come forward in the quantum corrections (in the sense that SUSY will not be broken at higher-order corrections as well).

To calculate the one-loop contribution to the effective potential, we will use the tadpole method \cite{Weinberg1}. For that we must calculate the superpropagators of the chiral superfields in the presence of a shift by a classical constant superfield. To get them we will use the techniques introduced in \cite{Prem}, which consists in to introduce two unconstrained scalar superfields (prepotentials)  $S$ and $\bar S$, from which the chiral and antichiral superfields are written as: 
\[\Phi=-\frac{1}{4}\bar{D}^2S; \quad\, \bar{\Phi}=-\frac{1}{4}D^2 \bar{S}.\]
The determination of the propagators of $S$ and $\bar S$ require the introduction of a gauge fixing term in the Lagragian, from which the final propagators of the chiral fields will not depend. With these substitutions, the action (\ref{dante1}) results in: 
\begin{equation}
{\mathcal S}=\int d^8z\left\{\bar{S}\Box\left[P_1\left(1-\frac{\Box}{M^2}\right)+\varepsilon (P_2+P_T)\right]S-\frac{1}{8}(mS\bar{D}^2S+\bar{m}\bar{S}D^2\bar{S})\right\} \ , \label{dante2}
\end{equation}
where
\begin{equation}\label{lc1}
P_1=\frac{D^2\bar{D}^2}{16\Box} \ , \ P_2=\frac{\bar{D}^2D^2}{16\Box} \ , \ P_T=\frac{\bar{D}_{\dot{\alpha}}D^2\bar{D}^{\dot{\alpha}}}{8\Box}
\end{equation} 
and $\varepsilon$ is the gauge fixing parameter. 

The propagators are obtained by inverting the quadratic terms on (\ref{dante2}), using the set of differential operators $P_1, P_2, P_T$, $P_+=\frac{1}{4\sqrt{\Box}}\bar{D}^2$ and $P_-=\frac{1}{4\sqrt{\Box}}D^2$, i.e.,
\begin{equation}
\Delta^{S\bar{S}}(z,z')=\frac{P_1\left(1-\frac{\Box}{M^2}\right)}{\Box\left(1-\frac{\Box}{M^2}\right)^2-\bar{m}P_{1}m}+\frac{1}{\varepsilon\Box}(P_2+P_T) \ .
\end{equation}
The superpropagators for $\Phi, \bar{\Phi}$ are obtained by acting with the appropriate $D^2, \bar{D}^2$ operators on the expression above, as
\[\Delta^{\Phi\bar{\Phi}}(z,z^\prime)=\frac{1}{16}\bar{D}^2D^{\prime 2}\Delta^{S\bar{S}}.\]

To derive the effective potential, we shift the chiral superfield by another constant chiral superfield ($\partial_\mu \Phi_c=0$),
\[\Phi\to\Phi+\Phi_c \ \ \ , \ \ \ \Phi_c=a+f\theta^2 \ .\]
The action of the shifted theory reads
\begin{eqnarray}\nonumber\label{miracle1}
{\mathcal S}&=&\int d^8z\bar{\Phi}\left(1-\frac{\Box}{M^2}\right)\Phi +\left(\int d^6z W''\Phi^2+h.c.\right)\\
&&+\int d^6z \left[\frac{g}{3!}\Phi^3+m\Phi_c\Phi+\frac{g}{2}\Phi_c^2\Phi\right]+\int d^8z\left[\bar{\Phi}_c\left(1-\frac{\Box}{M^2}\right)\Phi\right]+h.c. \ ,
\end{eqnarray}
where $W''(\Phi_c)=a+g\Phi_c=(m+ga)+gf\theta^2=\tilde{a}+\tilde{f}\theta^2.$ Performing the functional derivative with respect to the superfields $S$ and $\bar{S}$ and freely integrating by parts, one obtains that the superpropagators are related by the following equations:
\begin{eqnarray}
\Box\left[P_2\left(1-\frac{\Box}{M^2}\right)+\varepsilon (P_1+P_T)\right]\Delta^{\bar{S}\bar{S}}&=&\frac{1}{4}W''\bar{D}^2\Delta^{S\bar{S}} \ ,\\
\Box\left[P_1\left(1-\frac{\Box}{M^2}\right)+\varepsilon (P_2+P_T)\right]\Delta^{S\bar{S}}-\frac{1}{4}\bar{W''}D^2\Delta^{\bar{S}\bar{S}}&=&\delta^8(z-z').
\end{eqnarray}
The system above is immediately solved by
\begin{eqnarray}\nonumber
\Delta^{S\bar{S}}(z,z')&=&\left[\frac{P_1\left(1-\frac{\Box}{M^2}\right)}{\Box\left(1-\frac{\Box}{M^2}\right)^2-\bar{W''}P_1W''}+\frac{1}{\varepsilon \Box}(P_2+P_T)\right]\delta^8(z-z') \ , \\
\Delta^{\bar{S}\bar{S}}(z,z')&=&\frac{W''\bar{D}^2}{4\Box\left(1-\frac{\Box}{M^2}\right)}\Delta^{S\bar{S}}(z,z') \ .
\end{eqnarray}
Suppressing the $\varepsilon$ dependent term~\cite{Prem} (Landau gauge), $\Delta^{S\bar{S}}(z,z')$ can be expanded in a more convenient way, with its poles in explicit form, and we finally obtain
\begin{equation}\label{aj1}
\Delta^{S\bar{S}}=\frac{P_1}{\Box\left(1-\frac{\Box}{M^2}\right)^2-|\tilde{a}|^2}\delta^8(z-z')+e^{-i\theta\sigma\cdot\partial \bar{\theta}-i\theta^\prime\sigma\cdot\partial \bar{\theta}^\prime }\left[A\bar{\theta}^2\theta^{\prime 2}+B+C\bar{\theta}^2+E\theta^{\prime 2}\right]\delta^4(x-x') \ , 
\end{equation}
where
\begin{eqnarray}
A&=&\frac{|\tilde{f}|^2\left(1-\frac{\Box}{M^2}\right)^2}{\left\{\left[\Box\left(1-\frac{\Box}{M^2}\right)^2-|\tilde{a}|^2\right]^2-|\tilde{f}|^2\left(1-\frac{\Box}{M^2}\right)^2\right\}\left[\Box\left(1-\frac{\Box}{M^2}\right)^2-|\tilde{a}|^2\right]} \ , \nonumber\\
B&=&|\tilde{a}|^2\frac{A}{\Box^2\left(1-\frac{\Box}{M^2}\right)^2} \ , \nonumber\\
C&=&E^*=\frac{\tilde{a}\tilde{f}^*}{\Box\left[\left(\Box\left(1-\frac{\Box}{M^2}\right)^2-|\tilde{a}|^2\right)^2-|\tilde{f}|^2\left(1-\frac{\Box}{M^2}\right)^2\right]} \ .
\end{eqnarray}

The one-loop contribution to the effective potential can be calculated by integrating the superfield tadpole, shown in Fig. \ref{tadpole}. Its contribution is given by:
\begin{figure}
\begin{center}
\includegraphics[scale=.5]{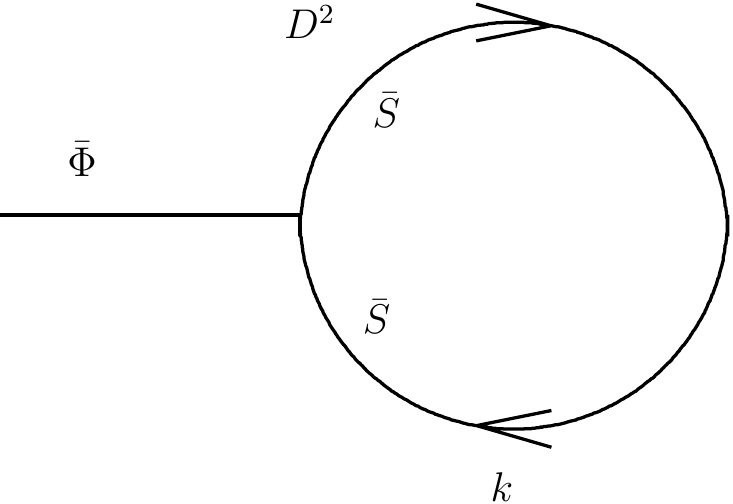}
\end{center}
\caption{Superfield tadpole \label{tadpole}}
\end{figure}
\begin{eqnarray}\label{chimaira3}\nonumber
i\Gamma_1&=&ig\int d^8z\bar{\Phi}(z)\left[-\frac{1}{4}D^2\Delta^{\bar{S}\bar{S}}(z,z')\right]_{z=z'}\\\nonumber
&=&i\int \frac{d^4 k}{(2\pi)^4} \frac{2g |\tilde{f}|^2X^2 \tilde{a}}{Y(Y^2-|\tilde{f}|^2 X^2)}\tilde{\bar A}(0)+i\int \frac{d^4 k}
{(2\pi)^4}\frac{X^2 g\tilde{f}}{Y^2-|\tilde{f}|^2 X^2}\tilde{\bar F}(0)\\
&=&-i\left(\frac{\partial V}{\partial a} \tilde{\bar A}(0)+\frac{\partial V}{\partial f}\tilde{\bar F}(0)\right)\ ,
\end{eqnarray}
where $X=1+\frac{k^2}{M^2}$ and $Y=k^2 X^2+|\tilde{a}|^2$. By integrating these equations we get the potential:
\begin{equation}\label{beer3}
V_1=-\frac{i}{2}\int\frac{d^4k}{(2\pi)^4}ln \left[1-\frac{|\tilde{f}|^2\left(1+\frac{k^2}{M^2}\right)^2}{\left[k^2\left(1+\frac{k^2}{M^2}\right)^2+|\tilde{a}|^2\right]^2}\right] \ .
\end{equation} 
By doing a Wick rotation to the Euclidian space ($k_0 =i k_4, k^2= \bold{k}^2 -k_0^2=\bold{k}^2+k_4^2=k_E^2 $), this potential can be written:
\begin{equation}\label{beer3}
V_1=\frac{1}{2}\int^{\Lambda}\frac{d^4k}{(2\pi)^4}ln \left[1-\frac{|\tilde{f}|^2\left(1+\frac{k^2}{M^2}\right)^2}{\left[k^2\left(1+\frac{k^2}{M^2}\right)^2+|\tilde{a}|^2\right]^2}\right] \ ,
\end{equation} 
where, in this expression and from now on, $k$ stands for the Euclidian momentum $k_E=(k_4,\bold{k})$ and $\Lambda (>> M^2)$ in the integral simbol represents an UV cut-off. It is easy to see that this expression is UV convergent, which means it is finite for $\Lambda/M \to \infty$.
If instead, we had considered $M^2>>\Lambda^2$, by expanding this expression in powers of $k^2/M^2$ we had obtain
\begin{eqnarray}\label{aj2}
V_1&=&\frac{1}{2}\int\frac{d^4k}{(2\pi)^4}\left\{ln\left(1-\frac{|\tilde{f}|^2}{(k^2+|\tilde{a}|^2)^2}\right)+O\left(\frac{1}{M^2}\right)\right\} .
\end{eqnarray}
As can be seen, the zero-order term is just the (UV logaritmic divergent) one-loop contribution to the effective potential \cite{grisaru1,Fogleman:1983hm} of the usual Wess-Zumino model.
The first correction, linear in $\frac{1}{M^2}$, and the followings are finite in the UV (which in this case means $\Lambda \to \infty$ but with $\Lambda/M<<1$.)

The poles in $k^2$ are given by
\begin{eqnarray}\nonumber
-m_1^2&=&-|\tilde{a}|^2-2\frac{|\tilde{a}|^4}{M^2}+\mathcal{O}(M^{-3}) \ ; \nonumber\\
-m_2^2&=&-|\tilde{a}|^2\pm|\tilde{f}|+(-2|\tilde{a}|^4-|\tilde{f}|^2+3|\tilde{a}|^2|\tilde{f}|)\frac{1}{M^2}+\mathcal{O}(M^{-3}) \ ; \nonumber\\
-m_3^2&=&-M^2\pm M|\tilde{a}|+\frac{|\tilde{a}|^2}{2}\mp\frac{5}{8}|\tilde{a}|^3\frac{1}{M}+\frac{|\tilde{a}|^4}{M^2}+\mathcal{O}(M^{-3}) \ ; \nonumber\\
-m_4^2&=&-M^2\mp M|\tilde{a}|+\frac{|\tilde{a}|^2}{2}-\frac{|\tilde{f}|}{2}\mp\frac{5|\tilde{a}|^4+|\tilde{f}|^2-6|\tilde{a}|^2|\tilde{f}|}{8|\tilde{a}|}M \nonumber\\
&&+\frac{1}{2}\left(2|\tilde{a}|^4+|\tilde{f}|^2-3|\tilde{a}|^2|\tilde{f}|\right)\frac{1}{M^2}+\mathcal{O}(M^{-3}) .
\end{eqnarray}
The quantities $m_1^2$ and $m_2^2$ are respectively the fermionic and bosonic masses corrected by terms of the order of $1/M^2$, $m_3^2$ and $m_4^2$ are the ghost masses introduced by the higher derivative term. By making $|\tilde{f}|^2=0$ , the fermionic and the bosonic masses become identicals, as well the ghost masses :
\begin{eqnarray}
-m^2_{1,2}&=&-|\tilde{a}|^2-2|\tilde{a}|^4M^2+\mathcal{O}(M^{-3})  \nonumber\\
-m^2_{3,4}&=&-M^2\pm M|\tilde{a}|^2\mp\frac{5}{8}|\tilde{a}|^3 M+\frac{|\tilde{a}|^4}{M^2}+\mathcal{O}(M^{-3}) \ .
\end{eqnarray}
Despite the existence of these states, we would stress the fact that the vacuum is only populated by the {\it physical} states. This can be seen from (\ref{rev1}), that the kinetic term for the field $A$ can be rewritten, using
\begin{equation}
B_1=\frac{1}{M^2}\Box A \ \ , \ \ B_2=-\frac{1}{M^2}\Box A+A \ , \label{red}
\end{equation}
as
\begin{equation}
L_{kin}(B_1,B_2)=B_2\Box B_2-B_1\left(\Box-M^2\right) B_1 \ , \label{red1}
\end{equation}
and one can separate the ghost from the physical particle using these redefinitions. In the vacuum, $\Box A=0\Rightarrow B_2=A \ , \ B_1=0$, as expected.


\section{One-Loop K\"{a}hlerian potential}

\begin{figure}
\begin{center}
\includegraphics[scale=.35]{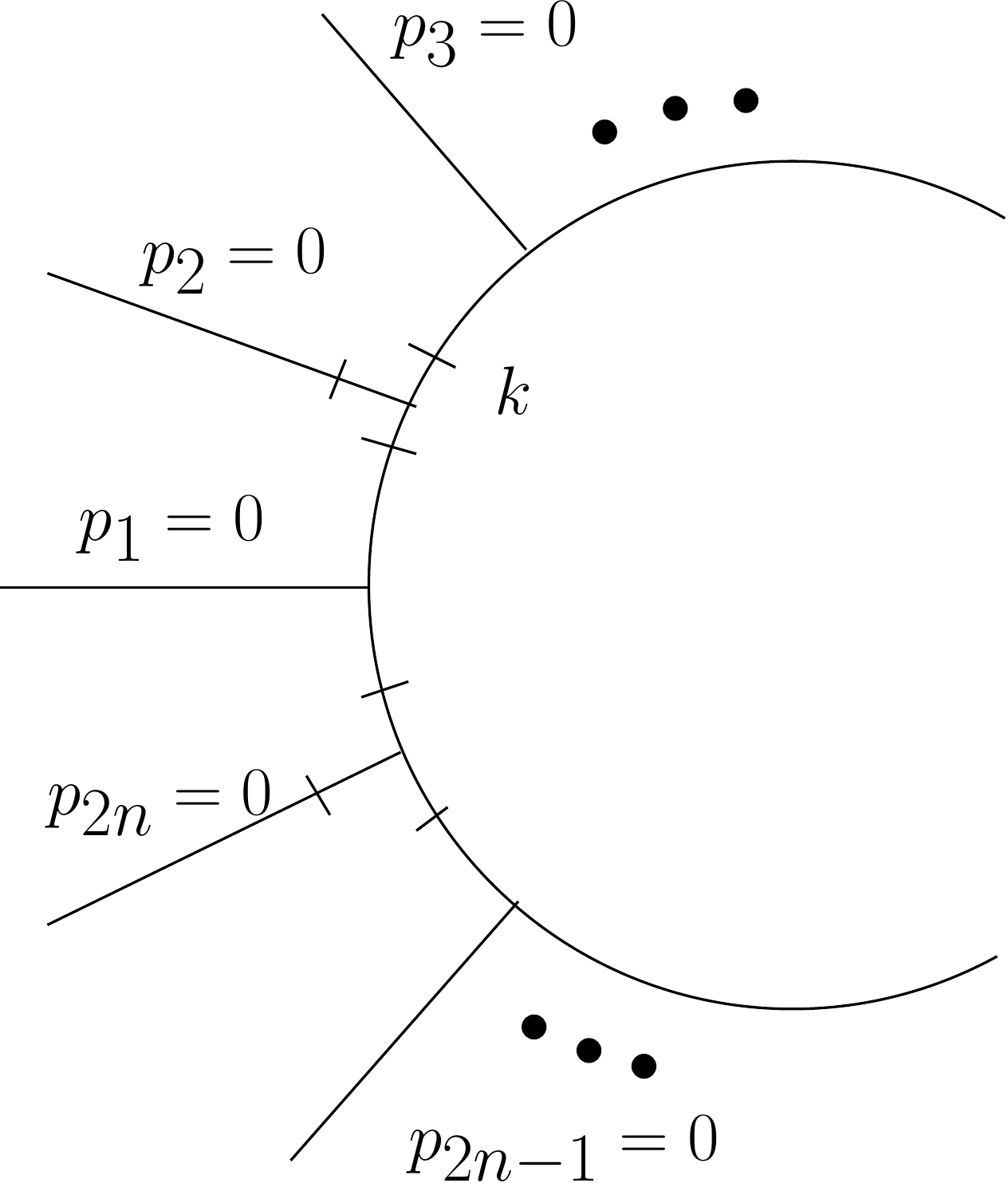}
\end{center}
\caption{Supergraphs contributing to the one-loop K\"{a}hlerian potential, massless case.}\label{tadpole34}
\end{figure}

We proceed with the calculation of the one-loop K\"{a}hlerian potential (where besides the condition $\partial_\mu \Phi_c=0$ we must impose, in the manipulations of the $D$-operator algebra, that $\bar{D}_{\dot{A}}\Phi_c=0$), as shown in Fig.~\ref{tadpole34}, for the massless theory, following \cite{Coleman,Gama:2011ws,Pickering:1996gt,Buchbinder:1994df}. We use again the unconstrained superfields $S, \bar{S}$ to obtain the propagators. We derive them from Eq.~(\ref{dante2}), 
\begin{eqnarray}\label{beer99}\nonumber
\Delta^{SS}&=&\frac{\bar{m}}{\Box\left(1-\frac{\Box}{M^2}\right)^2-m\bar{m}}\frac{P_{+}}{\sqrt{\Box}} \ \ \ , \ \ \ \Delta^{\bar{S}\bar{S}}=\frac{m}{\Box\left(1-\frac{\Box}{M^2}\right)^2-m\bar{m}}\frac{P_{-}}{\sqrt{\Box}} \ , \nonumber\\
\Delta^{S\bar{S}}&=&\frac{\left(1-\frac{\Box}{M^2}\right)}{\Box\left(1-\frac{\Box}{M^2}\right)^2-m\bar{m}}P_1+\frac{1}{\varepsilon\Box}(P_2+P_T) \ , \nonumber\\
\Delta^{\bar{S}S}&=&\frac{\left(1-\frac{\Box}{M^2}\right)}{\Box\left(1-\frac{\Box}{M^2}\right)^2-m\bar{m}}P_2+\frac{1}{\varepsilon\Box}(P_1+P_T) \ ,
\end{eqnarray} 
where $\varepsilon$ is a gauge fixing parameter. The same result can be obtained using Eq. (\ref{aj1}) in the limit $\Phi_c=0$.
With these propagators in hands, the one-loop K\"{a}hlerian potential for $m=\bar{m}=0$ is straightforwardly calculated. For the massless limit only supergraphs with the same number of $\Phi_c$ and $\bar\Phi_c$  external legs can contribute, and an $\frac{1}{2n}$ symmetry factor must be included, consequently one obtains:
\begin{eqnarray}\label{guns1}\nonumber
K^{(1)}&=&\frac{1}{2}\int\frac{d^8z}{\Box}\left.\displaystyle\sum_{n=1}^\infty\frac{1}{n}\left(\frac{g^2\Phi_c\bar\Phi_c}{\Box\left(1-\frac{\Box}{M^2}\right)^2}\right)^n\delta^4(x -x')\right|_{x=x'}\\
&=&-\frac{1}{2}\int\frac{d^8z}{\Box}\left. ln\left(1-\frac{g^2\Phi_c\bar\Phi_c}{\Box\left(1-\frac{\Box}{M^2}\right)^2}\right)\delta^4(x-x')\right|_{x=x'}.
\end{eqnarray}

Another way to derive (\ref{guns1}) is based on the methodology developed in \cite{Buchbinder:1994df}: from the action (\ref{dante1}) one can find the effective equations of motion for $\Phi,\bar{\Phi}$:
\begin{eqnarray}\label{koldb1}\nonumber
\frac{\delta {\mathcal S}}{\delta \Phi_c(z)}&=&-\frac{1}{4}\left(1-\frac{\Box}{M^2}\right)\bar{D}^2\bar\Phi_c(z)+W'[\Phi_c(z)],\\
\frac{\delta {\mathcal S}}{\delta \Phi_c(z)}&=&-\frac{1}{4}\left(1-\frac{\Box}{M^2}\right){D}^2\bar\Phi_c(z)+\bar{W}'[\bar\Phi_c(z)],
\end{eqnarray}
where $W[\Phi_c]=\frac{m}{2}\Phi_c^2+\frac{g}{3!}\Phi_c^3$. Now we can construct the matrix:
\begin{eqnarray}
\left(\begin{array}{cc} \frac{\delta^2 {\mathcal S}}{\delta \Phi_c(z)\delta\Phi_c(z')}&\frac{\delta^2 {\mathcal S}}{\delta \Phi_c(z)\bar\delta\Phi_c(z')} \\\frac{\delta^2 {\mathcal S}}{\delta \bar\Phi_c(z)\delta\Phi_c(z')}&\frac{\delta^2 {\mathcal S}}{\delta\bar \Phi_c(z)\bar\delta\Phi_c(z')}\end{array}\right)\!=\!\left(\begin{array}{cc}-\frac{1}{4}W''[\Phi_c]\bar{D}^2\delta^8(z,z')&\frac{1}{16}\left(1\!-\!\frac{\Box}{M^2}\right)\bar D^2 D^2\delta^8(z,z')\\\frac{1}{16}\left(1\!-\!\frac{\Box}{M^2}\right)D^2\bar{D}^2\delta^8(z,z')&-\frac{1}{4}\bar{W}''[\bar\Phi_c]D^2\delta^8(z,z')\end{array}\right) \ ,
\end{eqnarray}
so that the matrix superpropagator $G^{(W)}$ satisfies the equation
\begin{eqnarray}
-\left(\begin{array}{cc}W''[\Phi_c]&\left(1-\frac{\Box}{M^2}\right)\frac{1}{4}\bar D^2\\\left(1-\frac{\Box}{M^2}\right)\frac{1}{4}D^2&\bar{W}''[\bar\Phi_c]\end{array}\right)\left(\begin{array}{cc}G_{++}&G_{+-}\\G_{-+}&G_{--}\end{array}\right)=\left(\begin{array}{cc}\delta_{+}&0\\0&\delta_{-}\end{array}\right),
\end{eqnarray}
where $\delta_+=-\frac{1}{4}\bar{D}^2\delta^8(z-z')$ and $\delta_-=-\frac{1}{4}{D}^2\delta^8(z-z')$. For constant values of the background fields ( $\bar{D}_{\dot{A}}\Phi_c=0$) , the relevant part of this contribution is just
\begin{eqnarray}
G^{(W)}=-\frac{1}{\Box\left(1-\frac{\Box}{M^2}\right)^2-W''\bar{W}''}\left(\begin{array}{cc}\bar W''[\bar\Phi_c]&\left(1-\frac{\Box}{M^2}\right)\frac{1}{4}\bar D^2\\\left(1-\frac{\Box}{M^2}\right)\frac{1}{4}D^2&W''[\Phi_c]\end{array}\right)\left(\begin{array}{cc}\delta_{+}&0\\0&\delta_{-}\end{array}\right).
\end{eqnarray}
By changing the field definition to $\Phi_c\to i\Phi_c \ , \ \bar\Phi_c\to -i\bar\Phi_c$ \cite{Buchbinder}, whose only effect is to change the signal of the mass term in the action, we can write
\begin{eqnarray}
sTr\, ln[G^{(W)}]&=&\frac{1}{2}sTr\, ln[G^{(W)}G^{(-W)}] \nonumber\\ 
&=&\frac{1}{2}sTr\, \ln\left[\frac{1}{\Box\left(1-\frac{\Box}{M^2}\right)^2-W''\bar{W}''}\left(\begin{array}{cc}  1 &0 \\0& 1 \end{array}\right)\right] \nonumber\\
&=&\frac{1}{2}\int d^6z \left.  \ln\left(\frac{-1}{\Box\left(1-\frac{\Box}{M^2}\right)^2-W''\bar{W}''}\right)\delta_+(z,z')\right|_{z=z'} \nonumber\\
&&+\frac{1}{2}\int  d^6\bar{z}\left.   \ln\left(\frac{-1}{\Box\left(1-\frac{\Box}{M^2}\right)^2-W''\bar{W}''}\right)\delta_-(z,z')\right|_{z=z'} \nonumber\\
&=& \frac{1}{2}\left.\int d^8 z\frac{1}{\Box} \ln\left(\frac{-1}{\Box\left(1-\frac{\Box}{M^2}\right)^2-W''\bar{W}''}\right)\delta^8(z-z')\right|_{z=z'} \ , \label{koldpai1}
\end{eqnarray}
where we used the identity $\left. D^2\bar{D'}^2\delta^4(\theta-\theta')\right|_{\theta=\theta'}=16$ in the last line. Now (\ref{koldpai1}) can be cast in the well-known form:
\begin{equation}
K^{(1)}=-\frac{1}{2}\left.\int d^8z\frac{1}{\Box}\ln\left[1-\frac{W''[\Phi_c]\bar{W}''[\bar{\Phi_c}]}{\Box\left(1-\frac{\Box}{M^2}\right)^2}\right]\delta^{4}(x-x^{\prime})\right|_{x=x'} \ ,
\end{equation} 
where  $W''[\Phi_c]=m+g\Phi_c$ and $\bar{W}''[\bar{\Phi_c}]=m+g\bar{\Phi_c}$. As can be seen, for $m=0$ this is  equal to (\ref{guns1}). A similar result has been obtained in \cite{Flauger} using algebraic calculations.

Using that $\Phi_c=a+f\theta^2$, and so $W''=\tilde{a}+\tilde{f}\theta^2$, we obtain:
\begin{eqnarray}\nonumber
K^{(1)}&=&-\frac{1}{2}\int d^8z \frac{1}{\Box}\ln\left[1-\frac{(\tilde{a}+\tilde{f}\theta^2)(\tilde{a}^*+\tilde{f}^*\bar{\theta}^2)}{\Box\left(1-\frac{\Box}{M^2}\right)^2}\right]\delta^{4}(x-x^{\prime})\\\nonumber
&=&-\frac{1}{2}\int d^8z \frac{1}{\Box}\ln\left[\left(1-\frac{|\tilde{a}|^2}{\Box\left(1-\frac{\Box}{M^2}\right)^2}\right)\left(1-\frac{\tilde{a}\tilde{f}^*\bar{\theta}^2+\tilde{a}^*\tilde{f}\theta^2+|\tilde{f}|^2\theta^2\bar{\theta}^2}{\Box\left(1-\frac{\Box}{M^2}\right)^2-|\tilde{a}|^2}\right)\right]\delta^{4}(x-x^{\prime})\\\label{k1a}
&=&-\frac{1}{2}\int d^8z \frac{1}{\Box}\ln\left[1-\frac{\tilde{a}\tilde{f}^*\bar{\theta}^2+\tilde{a}^*\tilde{f}\theta^2+|\tilde{f}|^2\theta^2\bar{\theta}^2}{\Box\left(1-\frac{\Box}{M^2}\right)^2-|\tilde{a}|^2}\right]\delta^{4}(x-x^{\prime}),
\end{eqnarray}
since, from the second to the third line,  the integral of terms that have no dependence on $\theta, \bar{\theta}$ is null. So, we can write
\begin{eqnarray}
K^{(1)}&=&\frac{1}{2}\int \frac{d^4k}{(2\pi)^4}\frac{\left(1+\frac{k^2}{M^2}\right)^2}{\left[k^2\left(1+\frac{k^2}{M^2}\right)^2+|\tilde{a}|^2\right]^2}|\tilde{f}|^2.
\end{eqnarray}

This expression corresponds to  the first  order term in the expansion for the effective potential, Eq. (\ref{beer3}), in the auxiliary field  $|\tilde{f}|^2$ (this correspondence was proved for the O'Raifeartaigh model in \cite{div12}). Higher order terms in the expansion will be contributions to the  so called {\it auxiliary field potential }, $F(\bar{\Phi}, \Phi, D^2\Phi, \bar{D}^2\bar{\Phi})$ \cite{Buchbinder}.


\section{Concluding Remarks}

We succeeded in generalize the superfield tadpole method for higher-derivative theories. Using this methodology, we also carried out the explicit calculation of the one-loop effective potential in the supersymmetric higher-derivative Wess-Zumino model. Studying this result, we found that SUSY is not broken neither at the tree level nor at the one-loop level. The expression for the one-loop correction to the K\"{a}hlerian potential is also derived by two different methods, that is, by summation of Feynman supergraphs and the tadpole method. From this expression, we explicitly showed that the K\"{a}hler potential is the first order term in the expansion for the effective potential in powers of the auxiliary field. The main result of this paper consists in the development of a methodology, which seems useful for the  study of spontaneous SUSY breaking in more sophisticated theories, with at least two coupling constants. As a possible continuation of this study, we are planning to investigate some further aspects of this theory, which is interesting for both theoretical and  phenomenological points of view. 

In continuation to this study, we are investigating some further consequences of the introduction of higher derivatives in models wich exhibits spontaneous breaking of SUSY, looking for the relation of such breaking with this kind of operator. Another important task is to apply the methods developed here to models with more than one coupling constant, to investigate the possibility of dynamical symmetry breaking by the Coleman-Weinberg mechanism, and finally to study supersymmetric models with gauge fields, where we have the appearance of vector superfields. The calculations in this case turn out to be much more cumbersome and we have to pay attention to a nontrivial mixing between matter superfields and gauge potential superfields, wich imply important issues concerning the gauge fixing in superspace.


{\bf Acknowledgements.} This work was partially supported by Conselho
Nacional de Desenvolvimento Cient\'{\i}fico e Tecnol\'{o}gico (CNPq)
and Funda\c{c}\~{a}o de Amparo \`{a} Pesquisa do Estado de S\~{a}o
Paulo (FAPESP), Coordena\c{c}\~{a}o de Aperfei\c{c}oamento do Pessoal
do Nivel Superior (CAPES: AUX-PE-PROCAD 579/2008) and
CNPq/PRONEX/FAPESQ. The work by A. Yu. P. has been supported by the
CNPq project No. 303461/2009-8. C. R. Senise Jr. thanks Funda\c{c}\~{a}o de Amparo \`{a} Pesquisa do Estado de S\~{a}o
Paulo (FAPESP), project 2010/20797-2, for financial support.


\bibliographystyle{alpha}

\end{document}